\documentclass[conference]{IEEEtran}
\IEEEoverridecommandlockouts

\usepackage{cite}
\usepackage{amsmath,amssymb,amsfonts}
\usepackage{graphicx}
\usepackage{textcomp}
\usepackage{xcolor}
\usepackage{booktabs}
\usepackage{multirow}
\usepackage{enumitem}
\usepackage{url}
\usepackage{float}
\usepackage{stfloats}  
\usepackage[hidelinks]{hyperref}

\begin{document}

\title{REACT: A Conditioning Framework for\\
User-Adaptive sEMG Hand Pose Estimation}

\author{\IEEEauthorblockN{Eric Xie}
\IEEEauthorblockA{\textit{University of Toronto} \\
Toronto, Canada \\
ericx.xie@mail.utoronto.ca}
\and
\IEEEauthorblockN{Hei Shing Cheung}
\IEEEauthorblockA{\textit{University of Toronto} \\
Toronto, Canada \\
hayson.cheung@mail.utoronto.ca}
}

\maketitle

\begin{abstract}
Surface electromyography (sEMG) enables continuous hand pose estimation on wearable devices, but models trained on multi-user corpora degrade on unseen individuals due to inter-user variability in anatomy and electrode placement. We propose \textbf{REACT}, a lightweight conditioning framework that personalizes a frozen pretrained EMG-to-pose backbone at inference time using only a handful of calibration recordings. REACT learns a compact user embedding from calibration data and applies Feature-wise Linear Modulation (FiLM) to adapt the shared encoder's feature space, requiring no gradient updates at deployment. On the large-scale \textsc{emg2pose} benchmark, REACT improves over the state-of-the-art baseline across all three generalization splits in both regression and tracking modes, reducing angular error by up to 3.9\% with minimal parameter overhead and under 45 seconds of per-user calibration.
\end{abstract}


\begin{figure}[!h]
\centering
\includegraphics[width=\linewidth]{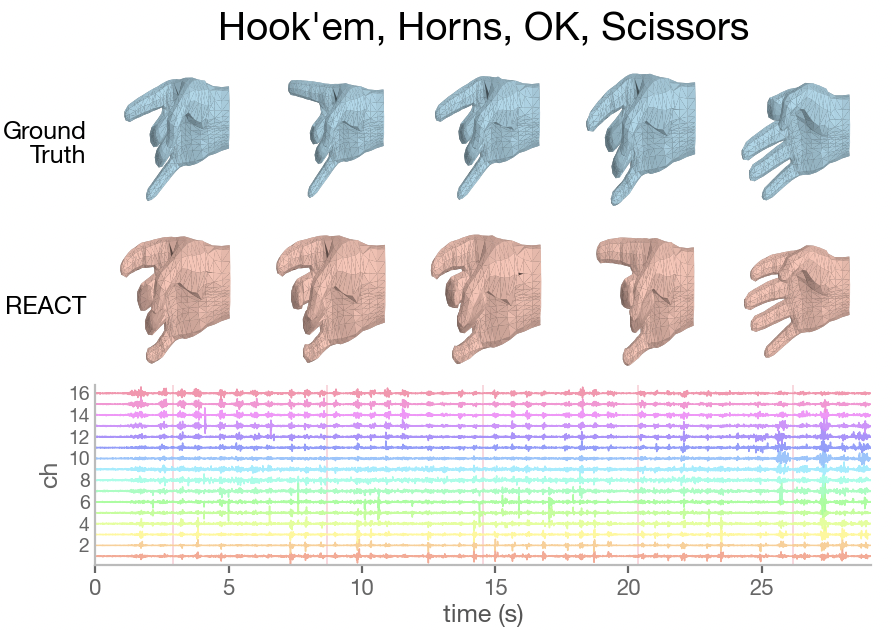}
\caption{\textbf{REACT reconstructs articulated hand pose from wristband sEMG.} For the \emph{Hook'em, Horns, OK, Scissors} gesture sequence, the \textbf{top} row shows the ground-truth mesh constructed from keypoints from a camera-based tracker, the \textbf{middle} row shows REACT's prediction, and the \textbf{bottom} strip shows the raw 16-channel EMG that is the only input at inference.}
\label{fig:teaser}
\end{figure}

\section{Introduction}
\label{sec:intro}

Continuous hand pose estimation from surface electromyography (sEMG) enables next-generation wearable interfaces for virtual and augmented reality, prosthetic control, and silent speech recognition~\cite{salter2024emg2pose, ctrlabs2024generic}. Unlike camera-based systems, sEMG-based approaches do not require line-of-sight and can operate in uncontrolled lighting, making them attractive for always-on wrist-worn devices. Recent large-scale efforts, notably \textsc{emg2pose}~\cite{salter2024emg2pose}, have collected hundreds of hours of synchronized EMG and pose data across nearly 200 users and demonstrated that deep temporal models can map multichannel sEMG to continuous hand joint angles with reasonable accuracy.

However, a persistent obstacle is \emph{inter-user variability}. Differences in muscle anatomy, subcutaneous fat, skin impedance, and electrode placement cause the same gesture to produce substantially different sEMG signatures across individuals~\cite{atzori2016electromyography, phinyomark2018feature}. Models trained on a large population degrade significantly when evaluated on held-out users or novel gesture stages~\cite{salter2024emg2pose}. Fine-tuning on per-user data is effective but expensive and impractical for consumer deployment. Meta-learning approaches such as MAML~\cite{finn2017maml} address few-shot adaptation but impose high computational overhead from second-order gradients.

We propose \textbf{REACT}, a framework that bridges the cross-user generalization gap through lightweight conditional adaptation. Given $k$ short calibration recordings from a new user ($\leq$45 seconds total), REACT generates a compact user embedding that drives Feature-wise Linear Modulation (FiLM)~\cite{perez2018film} of the encoder feature space, effectively personalizing a frozen pretrained backbone without modifying its weights at inference time. Our contributions are:

\begin{enumerate}[leftmargin=*,itemsep=2pt,topsep=2pt]
    \item \textbf{A modular user-adaptive conditioning pathway} that inserts between any frozen encoder-decoder pair without architectural changes to the backbone. The pathway combines characteristic convolutions, bidirectional GRU-based temporal pooling, transformer-based set aggregation, and FiLM projection to distill variable-length, variable-count calibration recordings into user-specific feature modulation.
    \item \textbf{Comprehensive evaluation} on the \textsc{emg2pose} benchmark across three orthogonal generalization axes and two prediction modes, demonstrating consistent improvements with real-time inferencing after brief calibration.
\end{enumerate}

\section{Related Work}
\label{sec:related}

\begin{figure*}[b]
\centering
\includegraphics[width=0.95\linewidth]{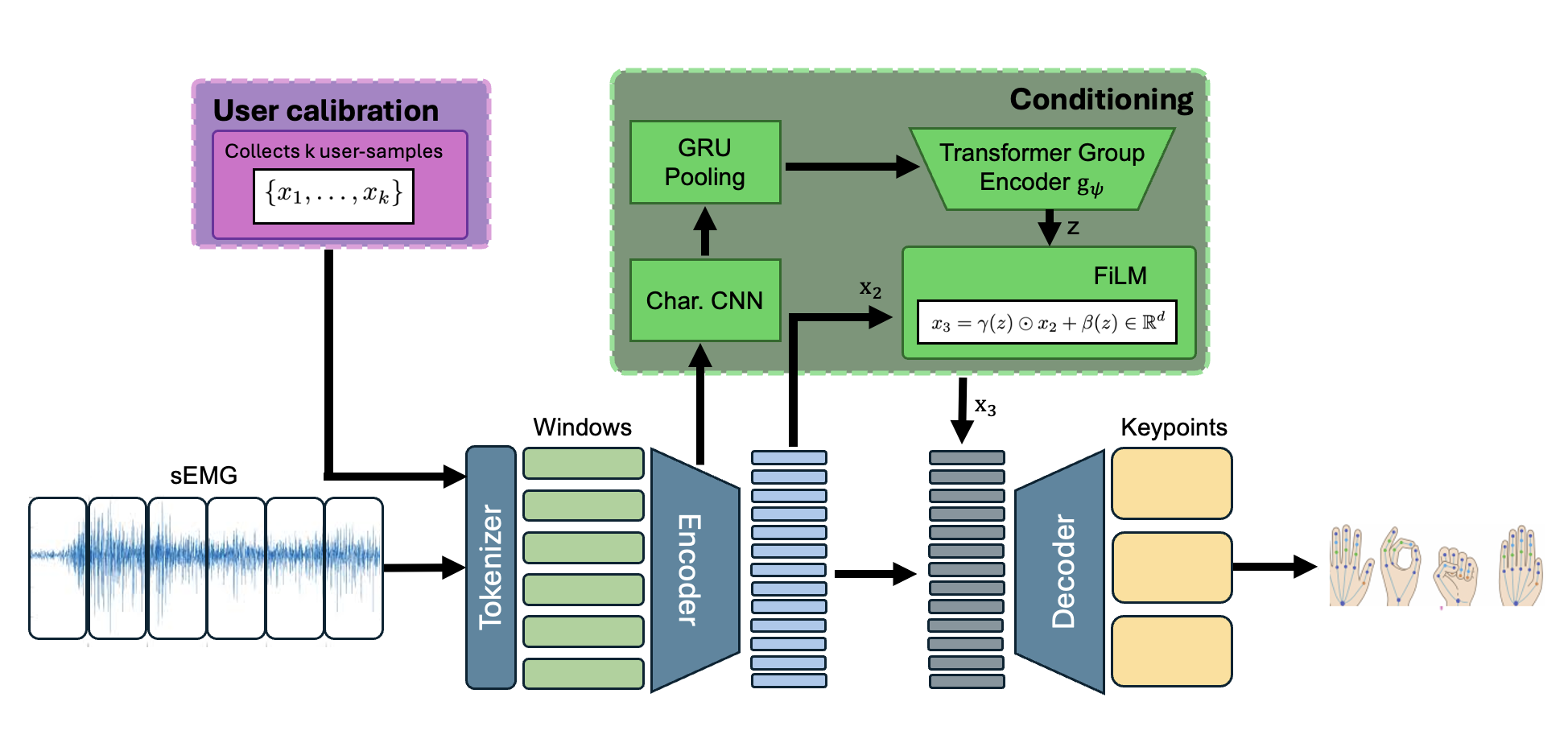}
\caption{REACT architecture. A pretrained TDS encoder maps raw EMG to temporal features. The conditioning pathway processes $k$ calibration recordings into a user embedding that drives a FiLM layer, which applies per-channel affine modulation $(\gamma, \beta)$ to the encoder features before the pretrained LSTM decoder.}
\label{fig:architecture}
\end{figure*}

\paragraph{EMG-based hand pose estimation}
Early approaches relied on hand-crafted time- and frequency-domain features fed to classifiers~\cite{phinyomark2012feature, atzori2016electromyography}. Deep learning methods have since dominated, with convolutional~\cite{atzori2016deep}, recurrent~\cite{ctrlabs2024generic}, and transformer-based~\cite{rahimian2021fs} architectures achieving strong within-user performance. The \textsc{emg2pose} benchmark~\cite{salter2024emg2pose} introduced a large-scale dataset with three supervised baselines (\textsc{vemg2pose}, NeuroPose, and SensingDynamics) evaluated across user, stage, and user+stage generalization splits. While \textsc{vemg2pose} achieves the best baseline accuracy, all models degrade substantially on unseen users, motivating user-adaptive solutions.

\paragraph{User adaptation and personalization}
Transfer learning and fine-tuning are widely used for EMG personalization~\cite{cote2019deep, du2017surface} but require gradient updates at deployment. Domain adaptation methods align source and target distributions~\cite{ganin2016domain} but typically assume access to labeled target data. More recently, CPEP~\cite{cui2025cpep} proposed contrastive pose-EMG pre-training for zero-shot gesture classification but did not address continuous pose regression or explicit user conditioning. Our approach differs by (i)~operating in the continuous pose regression setting, (ii)~performing test-time adaptation without gradient updates, and (iii)~using explicit user embeddings via FiLM rather than implicit alignment.

\paragraph{Feature-wise Linear Modulation}
FiLM~\cite{perez2018film} applies learned affine transformations $\gamma \odot \mathbf{f} + \beta$ conditioned on auxiliary information, originally proposed for visual question answering. It has since been adopted for style transfer~\cite{dumoulin2017learned}, few-shot learning~\cite{oreshkin2018tadam}, multi-task learning~\cite{strezoski2019many}, and speaker-conditioned keyword spotting in speech processing~\cite{labrador2025keyword}. We adapt FiLM for user-specific conditioning in the biosignal domain, where it provides a natural mechanism for modulating shared representations with per-user characteristics.

\paragraph{Temporal pooling and set aggregation}
Aggregating variable-length sequences and variable-size sets into fixed representations are fundamental challenges in our pipeline. Gated Recurrent Units (GRUs)~\cite{cho2014gru} provide effective temporal summarization by maintaining a hidden state across variable-length sequences; we use a bidirectional GRU to pool per-recording features, as it captures ordering information that attention-based pooling discards. For aggregating multiple recordings, Set Transformer~\cite{lee2019set} introduced attention-based pooling with learnable query tokens; our transformer group encoder draws on this idea, using a learnable \texttt{[CLS]}-style query to aggregate $k$ recording-level embeddings into a single user representation.

\section{Method}
\label{sec:method}

\subsection{Problem formulation}
\label{sec:formulation}

Let $\mathbf{x} \in \mathbb{R}^{16 \times L}$ denote a 16-channel sEMG recording of length $L$ at 2\,kHz, and $\mathbf{y} \in \mathbb{R}^{20 \times T}$ the corresponding 20-DOF hand joint angles at $T = 50$\,Hz. For user $u$, we additionally have access to a calibration set $\mathcal{C}_u = \{\mathbf{c}_1^u, \ldots, \mathbf{c}_k^u\}$ of $k$ short recordings from other sessions of the same user. The goal is to learn a mapping $f(\mathbf{x}; \mathcal{C}_u) \rightarrow \hat{\mathbf{y}}$ that adapts to user $u$ at inference time using only $\mathcal{C}_u$, without gradient updates.

\subsection{Architecture overview}
\label{sec:architecture}

REACT inserts a conditioning pathway between the pretrained \textsc{vemg2pose} encoder and decoder, as shown in Figure~\ref{fig:architecture}.

\subsection{Pretrained backbone}
\label{sec:backbone}

We build on the \textsc{vemg2pose} model~\cite{salter2024emg2pose}, a TDS convolutional encoder~\cite{hannun2019sequence} paired with a two-layer LSTM decoder. The encoder maps raw sEMG to temporal features; the decoder supports both regression (absolute joint angle prediction) and tracking (velocity integration from an initial pose) modes. Full architectural details are in Appendix~\ref{app:implementation}.

\subsection{Conditioning}
\label{sec:conditioning}

Each calibration recording is processed through four stages to produce user-specific modulation of the encoder features:

\begin{enumerate}[leftmargin=*,itemsep=2pt,topsep=2pt]
    \item A \emph{Characteristic CNN} that extracts user-specific features from the shared encoder output.
    \item A \emph{Bidirectional GRU}~\cite{cho2014gru} that pools each recording's temporal dimension into a fixed-size vector, capturing ordering information that attention pooling discards.
    \item A \emph{Transformer Group Encoder}~\cite{lee2019set} that aggregates the $k$ per-recording vectors into a single user embedding $\mathbf{z}$. When $k{=}0$ (no calibration), a learnable fallback embedding $\mathbf{z}_0$ is used.
    \item A \emph{FiLM projection} that maps the embedding to $(\gamma, \beta) \in \mathbb{R}^{64}$, applied channel-wise as $\gamma \odot \mathbf{f} + \beta$ to the encoder output, allowing user-specific rescaling without altering the pretrained feature structure.
\end{enumerate}

For stable training, $\gamma$ is initialized to $\mathbf{1}$ and $\beta$ to $\mathbf{0}$, ensuring the model replicates the pretrained baseline at initialization. Architectural hyperparameters are provided in Appendix~\ref{app:implementation}.

\subsection{Training procedure}
\label{sec:training}

Training proceeds in two phases to balance adaptation learning and backbone preservation.

\paragraph{Phase 1: Adaptation module training (frozen backbone)}
The encoder and decoder are frozen; only the conditioning pathway is trained with AdamW~\cite{loshchilov2019decoupled} and cosine annealing for 2 epochs. At each iteration, $k \sim \text{Uniform}(0, 30)$ calibration recordings are sampled from other sessions of the same user. Calibration recordings are pre-encoded offline to avoid redundant computation.

\paragraph{Phase 2: Joint fine-tuning (unfrozen backbone)}
The encoder and decoder are unfrozen with layer-wise learning rate decay for 2 additional epochs. This prevents catastrophic forgetting of pretrained representations while allowing the backbone to co-adapt with the conditioning module.

\paragraph{Loss function}
The combined training objective is:
\begin{equation}
    \mathcal{L} = \underbrace{\frac{1}{N}\sum_{i} |\hat{y}_i - y_i|}_{\text{Angular MAE}} \;+\; \lambda \underbrace{\frac{1}{N}\sum_{i} \|\hat{\mathbf{p}}_i - \mathbf{p}_i\|_2}_{\text{Fingertip distance}},
    \label{eq:loss}
\end{equation}
where $\hat{y}_i, y_i$ are predicted and ground-truth joint angles (in radians), $\hat{\mathbf{p}}_i, \mathbf{p}_i$ are 3-D fingertip positions derived via forward kinematics, and $\lambda = 0.01$. Invalid labels (inverse-kinematics failures) are masked during loss computation.

\paragraph{Data augmentation}
Channel rotation augmentation simulates electrode placement variability by cyclically shifting the EMG channels~\cite{salter2024emg2pose}. Instance normalization is applied per channel per recording.

\section{Experiments}
\label{sec:experiments}

\subsection{Dataset}
\label{sec:dataset}

We evaluate on the \textsc{emg2pose} benchmark~\cite{salter2024emg2pose}, which provides synchronized sEMG and hand pose labels. Table~\ref{tab:dataset} summarizes the dataset statistics.

\begin{table}[h]
\centering
\caption{\textsc{emg2pose} dataset statistics.}
\label{tab:dataset}
\small
\begin{tabular}{lccc}
\toprule
\textbf{Split} & \textbf{Users} & \textbf{Recordings} & \textbf{Hours} \\
\midrule
Train & 158 & 17{,}136 & 288.9 \\
Validation & 15 & 1{,}950 & 32.6 \\
Test & 178$^\dagger$ & 6{,}167 & 102.7 \\
\midrule
Total & 193 & 25{,}253 & 424.2 \\
\bottomrule
\end{tabular}
\\[3pt]
{\footnotesize $^\dagger$Test includes 20 held-out users plus 158 training users evaluated on unseen gesture stages.}
\end{table}

\subsection{Evaluation protocol}
\label{sec:eval_protocol}

Following~\cite{salter2024emg2pose}, we evaluate across three orthogonal generalization axes:
\begin{itemize}[leftmargin=*,itemsep=1pt,topsep=2pt]
    \item \textbf{User split:} Held-out users, seen gesture stages. Tests adaptation to new users.
    \item \textbf{Stage split:} Seen users, held-out gesture stages. Tests generalization to novel movements.
    \item \textbf{User+Stage split:} Both held out. The hardest generalization setting.
\end{itemize}

We report two metrics aligned with the benchmark: \textbf{angular MAE} (degrees) and \textbf{landmark distance} (mm). Both modes of the baseline are evaluated: \emph{regression} (absolute joint angle prediction) and \emph{tracking} (velocity integration from an initial pose). Our baseline is the unmodified pretrained \textsc{vemg2pose}~\cite{salter2024emg2pose}, the current state-of-the-art on \textsc{emg2pose}. We use the publicly released checkpoint without retraining, ensuring a fair comparison: REACT builds directly on top of this backbone, so any improvement must come from the conditioning module.

\subsection{Quantitative results}
\label{sec:main_results}

Table~\ref{tab:results} presents results at calibration budget $k{=}15$ across all three generalization splits and both prediction modes. REACT outperforms \textsc{vemg2pose} on \textbf{all metrics in both regression and tracking}. The largest gains appear on the Stage split for regression (MAE: $15.2 \to 14.6^\circ$, 3.9\% reduction) and the User split for tracking (MAE: $7.7 \to 7.4^\circ$, 3.9\% reduction), where user-specific conditioning most directly addresses the domain gap.

\begin{table*}[t]
\centering
\caption{\textbf{Results at $k{=}15$ calibration recordings.} Angular error (degrees, $\downarrow$) and landmark distance (mm, $\downarrow$). Bold indicates best. REACT improves on all splits and modes.}
\label{tab:results}
\begin{tabular}{llcccc}
\toprule
& & \multicolumn{2}{c}{\textbf{Regression}} & \multicolumn{2}{c}{\textbf{Tracking}} \\
\cmidrule(lr){3-4} \cmidrule(lr){5-6}
\textbf{Split} & \textbf{Model} & \textbf{Err ($^\circ$)} & \textbf{Dist (mm)} & \textbf{Err ($^\circ$)} & \textbf{Dist (mm)} \\
\midrule
\multirow{2}{*}{User}
  & \textsc{vemg2pose}    & $12.2$ & $15.8$ & $7.7$  & $10.3$ \\
  & \textbf{REACT (ours)} & $\mathbf{12.0}$ & $\mathbf{15.6}$ & $\mathbf{7.4}$  & $\mathbf{10.0}$ \\
\midrule
\multirow{2}{*}{Stage}
  & \textsc{vemg2pose}    & $15.2$ & $20.4$ & $11.2$ & $15.2$ \\
  & \textbf{REACT (ours)} & $\mathbf{14.6}$ & $\mathbf{19.9}$ & $\mathbf{10.9}$ & $\mathbf{14.9}$ \\
\midrule
\multirow{2}{*}{User+Stage}
  & \textsc{vemg2pose}    & $15.8$ & $21.6$ & $11.0$ & $15.4$ \\
  & \textbf{REACT (ours)} & $\mathbf{15.3}$ & $\mathbf{21.2}$ & $\mathbf{10.8}$ & $\mathbf{15.3}$ \\
\bottomrule
\end{tabular}
\end{table*}

Tracking mode consistently outperforms regression across all models ($7.4^\circ$ vs.\ $12.0^\circ$ for REACT on the User split), since velocity integration with a known initial pose provides a strong prior that constrains the solution space. This advantage comes at the cost of requiring the initial pose, which may not always be available in deployment.

\subsection{Qualitative results}
\label{sec:qualitative}

\begin{figure*}[t]
\centering
\includegraphics[width=\linewidth]{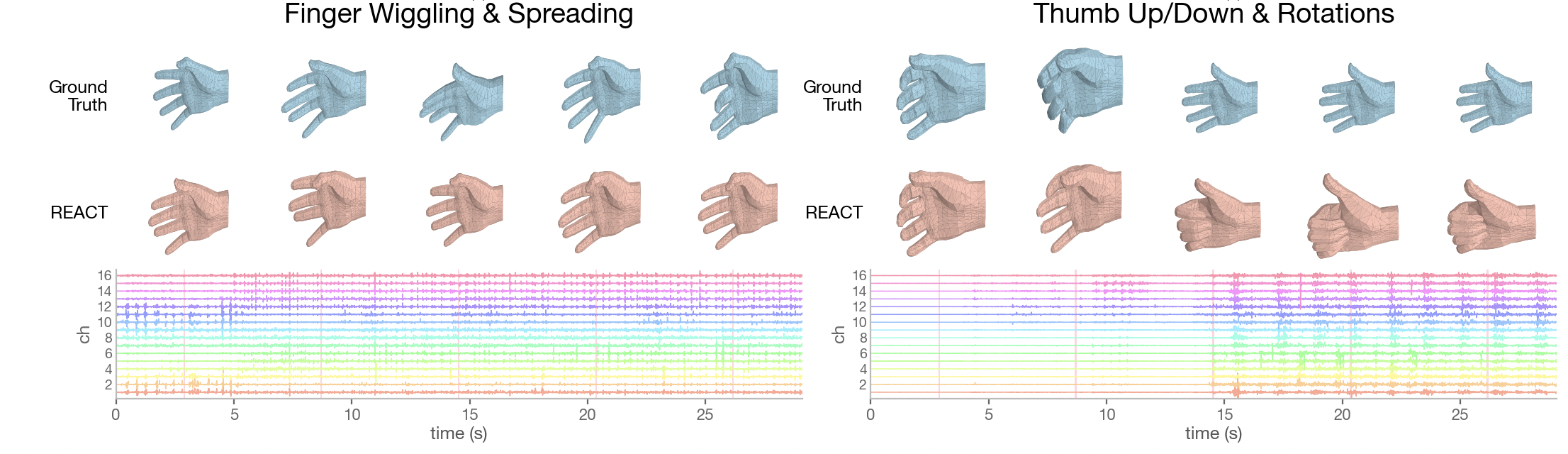}
\caption{\textbf{Qualitative reconstructions for two held-out gesture sequences} (left: \emph{Finger Wiggling \& Spreading}; right: \emph{Thumb Up/Down \& Rotations}). For each sequence, the \textbf{top} row shows the ground-truth mesh and keypoints from a camera-based tracker, the \textbf{middle} row shows REACT's prediction at the same timesteps, and the \textbf{bottom} strip shows the raw 16-channel EMG input. All frames are from a held-out user.}
\label{fig:qualitative}
\end{figure*}

REACT's pose reconstructions (Figure~\ref{fig:qualitative}) track the ground truth closely across diverse gestures, from finger wiggling and spreading to thumb rotations, producing anatomically plausible meshes throughout the sequences. Per-finger analysis reveals that REACT's improvement is most pronounced for the pinky ($11.4^\circ \to 11.1^\circ$ in tracking) and middle fingers, which exhibit the highest inter-user variability. The thumb, which has the most constrained range of motion, shows the smallest baseline error and thus the smallest absolute improvement.

\section{Discussion}
\label{sec:discussion}

\paragraph{Why does FiLM conditioning work?}
We hypothesize that inter-user variability manifests primarily as channel-wise scale and offset differences in the encoder feature space, arising from differences in muscle activation amplitudes and electrode impedance. FiLM's affine transformation is well-suited to correct these per-channel distributional shifts. The identity initialization ($\gamma{=}1, \beta{=}0$) ensures that REACT replicates the pretrained baseline before learning, providing a conservative adaptation that cannot harm pre-existing performance at initialization. Despite its negligible parameter cost, the FiLM layer produces consistent improvements, suggesting that the pretrained encoder features are already expressive; they merely need user-specific rescaling to account for individual signal characteristics.

\paragraph{A backbone-agnostic add-on}
REACT is a conditioning module, not a standalone estimator: it reshapes the feature space of whatever encoder--decoder backbone it wraps rather than predicting pose on its own. Its accuracy ceiling is therefore inherited from that backbone. REACT can sharpen a strong model but cannot rescue a weak one. The upside is flexibility: because the conditioning pathway makes no architectural assumptions about the backbone, it can be attached to any future state-of-the-art EMG-to-pose model to eke out additional accuracy.

\paragraph{Regression vs.\ tracking}
REACT's improvements are larger in tracking mode (up to 3.9\% angular MAE reduction) than regression mode (up to 3.2\%). We attribute this to tracking's recurrent velocity integration, where small per-timestep improvements compound over the prediction horizon. Regression predicts each frame more independently, limiting the compounding benefit of conditioning.

\paragraph{Limitations}
Several limitations warrant discussion: (i)~Our improvements, while consistent, are moderate (2 to 4\%); more aggressive adaptation (e.g., adding FiLM to multiple encoder layers) may yield larger gains at the cost of increased complexity. (ii)~We evaluate only on the \textsc{emg2pose} dataset; generalization to other EMG devices, electrode configurations, or clinical populations remains untested. (iii)~The current design applies FiLM at a single point in the architecture; multi-scale conditioning may capture richer user characteristics.

\paragraph{Beyond EMG}
The principle behind REACT is domain-agnostic: it addresses cross-user adaptation, where a population-trained model must be calibrated to an individual whose signal statistics differ from the training distribution. The same bottleneck arises in EEG, wearable sensing, and biometric monitoring. REACT offers a trained conditioning module as a potential solution to overcome this challenge. We see this as a promising direction for any domain where user calibration is a recurring obstacle.

\section{Conclusion}
\label{sec:conclusion}

We presented REACT, a lightweight conditioning framework for EMG-to-pose estimation that bridges the cross-user generalization gap through Feature-wise Linear Modulation. By inserting a modular conditioning pathway between a frozen pretrained encoder and decoder, REACT personalizes real-time predictions using only a handful of calibration recordings. On the \textsc{emg2pose} benchmark, REACT consistently improves angular error and landmark distance across all generalization splits in both regression and tracking modes. Our results demonstrate that conditional adaptation is a practical strategy for deploying EMG-based hand pose systems across diverse user populations. More broadly, by framing personalization as gradient-free conditioning of a frozen backbone, REACT offers a reusable recipe for any domain facing user calibration or cross-user adaptation challenges.

For future work, promising directions include: (i)~multi-scale FiLM conditioning at multiple encoder layers, (ii)~contrastive pre-training of the user encoder for richer embeddings, (iii)~online adaptation during continuous use, (iv)~evaluation on clinical populations with neuromotor conditions, and (v)~transferring the conditioning pathway to non-EMG modalities that share the cross-user calibration problem, such as EEG, IMU-based wearable sensing, and biometric monitoring.

\section*{Acknowledgment}

We thank Meta Reality Labs for releasing the \textsc{emg2pose} dataset and pretrained baselines. Compute resources were provided by Modal (H100 GPU cloud). This work was conducted as part of APS360 at the University of Toronto.


\appendices

\section{Implementation Details}
\label{app:implementation}

\paragraph{Encoder architecture}
The TDS encoder consists of two strided 1-D convolution blocks followed by two TDS stages:
\begin{itemize}[leftmargin=*,itemsep=1pt]
    \item Conv1dBlock: $16 \rightarrow 256$ channels, kernel size 11, stride 5
    \item Conv1dBlock: $256 \rightarrow 256$ channels, kernel size 5, stride 2
    \item TDS Stage 1: 2 blocks, $\text{in\_channels}{=}256$, $\text{channels}{=}16$, $\text{feature\_width}{=}16$, $\text{kernel}{=}9$
    \item TDS Stage 2: 2 blocks, same config with $\text{kernel}{=}5$, $\text{out\_channels}{=}64$
\end{itemize}
Left context: 1,790 samples at 2\,kHz ($\approx$0.9\,s). Input windows: 11,790 samples (10,000 content + 1,790 context), strided by 2,000 samples (1\,s).

\paragraph{Decoder architecture}
The LSTM decoder is a 2-layer LSTM with hidden size 512:
\begin{itemize}[leftmargin=*,itemsep=1pt]
    \item Regression mode: $\text{in\_channels}{=}84$ (64 features + 20 state), $\text{out\_channels}{=}40$ (position + velocity)
    \item Tracking mode: $\text{in\_channels}{=}84$, $\text{out\_channels}{=}20$ (velocity only)
    \item Output scale: 0.01; operating frequency: 50\,Hz
\end{itemize}
In regression mode, the first 500 samples at 2\,kHz ($\equiv$12.5 steps at 50\,Hz) use direct position output, then switch to velocity integration.

\paragraph{User encoder details}
The characteristic CNN uses kernel size 3, layer normalization, GELU activations, and dropout. The bidirectional GRU has 3 layers, hidden size 64, and dropout 0.1. The transformer group encoder uses 3 layers, 4 attention heads, and hidden dimension 128, with sinusoidal positional encodings supporting up to 32 calibration recordings. The FiLM projection MLP uses LayerNorm and GELU activations, with $\gamma$ initialized to $\mathbf{1}$ and $\beta$ to $\mathbf{0}$.

\paragraph{Parameter counts}
The conditioning pathway adds $\sim$890K parameters, inserted between the pretrained \textsc{vemg2pose} encoder and decoder ($\sim$5.9M parameters combined). The conditioning pathway thus accounts for roughly 13\% of the backbone size.

\paragraph{Training hyperparameters}
Phase~1: AdamW with lr $= 10^{-3}$, weight decay $= 10^{-4}$, cosine annealing, and 500-step linear warmup. Phase~2: base lr $= 10^{-4}$ with decay factor 0.5 per layer depth.

\paragraph{Training infrastructure}
All experiments were conducted on a single NVIDIA H100 GPU via Modal cloud computing. Batch size: 128. Mixed-precision training (AMP) was enabled. Total training time: $\sim$4 hours for the full two-phase procedure.

\section{Decoder Operating Modes}
\label{app:modes}

The \textsc{vemg2pose} decoder supports two complementary modes:

\textbf{Regression mode} predicts absolute joint angles from each EMG window independently. The decoder outputs a 40-dimensional vector (20 positions + 20 velocities) at each 50\,Hz timestep. For the first 12.5 steps ($= 500$ samples at 2\,kHz), the position output is used directly; thereafter, velocity is integrated: $\hat{\mathbf{y}}_t = \hat{\mathbf{y}}_{t-1} + \hat{\mathbf{v}}_t$.

\textbf{Tracking mode} requires an initial pose (from ground truth) and predicts velocities only (20-dimensional output). Pose is recovered via pure integration: $\hat{\mathbf{y}}_t = \hat{\mathbf{y}}_{t-1} + \hat{\mathbf{v}}_t$ for all timesteps. This mode typically achieves lower absolute error since it benefits from the ground-truth anchor, but requires the initial pose to be known.

\end{document}